\documentclass{aa}  

\usepackage{graphicx}
\usepackage{txfonts}
\usepackage{makecell}
\usepackage{subfiles}
\usepackage{siunitx}
\usepackage[normalem]{ulem}
\usepackage{threeparttable}

\DeclareMathAlphabet{\mathcal}{OMS}{cmsy}{m}{n}

\newcommand\teff{\ensuremath{T_{\textup{eff}}}\xspace}
\newcommand\logg{\ensuremath{\log g}\xspace}
\newcommand\muHz{\ensuremath{\mu \mathrm{Hz}}\xspace}
\newcommand\numax{\ensuremath{\nu_{\textup{max}}}\xspace}
\newcommand\dnu{\ensuremath{\Delta\nu}\xspace}

\newcommand\FeH{\ensuremath{[\mathrm{Fe}/\mathrm{H}]}\xspace}

\newcommand\kepler{\textit{Kepler}\xspace}

\begin{document} 
    \title{Granulation signatures in 3D hydrodynamical simulations: evaluating background model performance using a Bayesian nested sampling framework}

   \author{J. R. Larsen\inst{1}\fnmsep\thanks{E-mail: jensrl@phys.au.dk}
            \and
            M. S. Lundkvist\inst{1}
            \and
            G. R. Davies\inst{2}
            \and
            M. B. Nielsen\inst{2}
            \and
            H.-G. Ludwig\inst {3}
            \and 
            Y. Zhou\inst{4,5,1}
            \and
            L. F. Rodríguez Díaz\inst{1}
            \and
            H. Kjeldsen\inst{1,6}
        }
        
   \institute{Stellar Astrophysics Centre (SAC), Dept. of Physics and Astronomy, Aarhus University,
              Ny Munkegade 120, 8000 Aarhus C, Denmark 
              \and
              School of Physics and Astronomy, University of Birmingham, Edgbaston B15 2TT, United Kingdom 
              \and
              Center for Astronomy (ZAH/LSW), Heidelberg University, Königstuhl 12, 69117 Heidelberg, Germany
              \and
              Rosseland Centre for Solar Physics, Institute of Theoretical Astrophysics, University of Oslo, P.O. Box 1029, Blindern, NO-0315 Oslo, Norway
              \and
              School of Physical and Chemical Sciences --- Te Kura Mat{\=u}, University of Canterbury, Private Bag 4800, Christchurch 8140, Aotearoa, New Zealand              
              \and
              Aarhus Space Centre (SpaCe), Dept. of Physics and Astronomy, Aarhus University, Ny Munkegade 120, 8000 Aarhus C, Denmark 
            }
   \date{Received 27 May 2025; Accepted 14 July 2025}

  \abstract
   {Understanding the granulation background signal is of vital importance when interpreting the asteroseismic diagnostics of solar-like oscillators. Various descriptions exist in the literature for modelling the surface manifestation of convection, the choice of which affects our interpretations. 
   }
   {We aim to evaluate the performance of and preference for various granulation background models for a suite of 3D hydrodynamical simulations across the Hertzsprung-Russel diagram, thereby expanding the number of simulations and coverage of parameter space for which such investigations have been made.
   }
   {We take a statistical approach by considering the granulation signatures in power density spectra of 3D hydrodynamical simulations of convection, where no biases or systematics of observational origin are present. To properly contrast the performance of the background models, we develop a Bayesian nested-sampling framework for model inference and comparison. This framework is subsequently extended to real stellar data using the solar analogue KIC 8006161 (\textit{Doris}) and the Sun.
   }
   {We find that multi-component models are consistently preferred over a single-component model, with each tested multi-component model demonstrating merit in specific cases. This occurs for simulations with no magnetic activity, thus ruling out stellar faculae as the sole source of the second granulation component. Similar to a previous study, we find that a hybrid model with a single overall amplitude and two characteristic frequencies performs well for numerous simulations. Additionally, a tentative third granulation component beyond the value of \numax is seen for some simulations, but its potential presence in observations requires further studies.
   }
  {Studying the granulation signatures in these simulations paves the way to studying real stars with accurate granulation models. This deeper understanding of the granulation signal may lead to complementary methods to existing algorithms for determining stellar parameters, aiming for an independent radius estimate for stars where oscillations are not observable. With the steadily increasing catalogue of stars that possess extended timeseries, supplied by TESS and the upcoming PLATO mission, exploring this avenue could be highly beneficial.
  }

   \titlerunning{Granulation signatures in 3D hydrodynamical simulations}  
   \maketitle

\section{Introduction}\label{sec:Intro}
The modelling of turbulent convection in stars is a longstanding challenge in modern astrophysics. Not only its treatment in stellar evolution codes through the mixing length theory \citep{VitenseMLT58,
Kippenhahn13,Trampedach14,LiY2024} but also the interpretation of the surface signatures of convection, granulation, deserves careful consideration. The convective motions are sensitive to the density profile of the outer regions of a solar-like star, and any star with a convective envelope will display granulation. Several studies have investigated this phenomenon observationally \citep{Mathur11, Kallinger2014} and, in line with the theoretical predictions, find that the amplitudes and timescales of the granulation scale with the frequency of maximum oscillation power, $\numax$ \citep{Kjeldsen11, Chaplin11b}. The scaling of \numax with stellar evolution, like that of convective motions, depends on fundamental stellar parameters, following $\numax\propto g/\sqrt{\teff}$. This suggests a link between granulation-related signals and global stellar properties, a relationship that is exploited by algorithms such as Flicker and FliPer to estimate surface gravities \citep{Bastien16,Bugnet18}.

In this work, we consider the quasi-periodic brightness fluctuations caused by the granulation -- which occur over a wide range of amplitudes and timescales -- from a theoretical standpoint using 3D hydrodynamical simulations of convection (hereafter simply simulations). Several previous studies have taken this theoretical approach when investigating the signatures of granulation (for an extensive comparison of simulations and observations, see \citealt{Zhou21}). \citet{Mathur11} were the first to consider the power density spectra (PDS) of such simulations, finding that the trends broadly reproduced the observed scaling relations for granulation amplitudes and timescales in their red giant sample, albeit with notable systematic discrepancies. \citet{Samadi13a, Samadi13b} used a theoretical model for the PDS of the simulations to support the observed scaling with \numax, however also noting the importance of the Mach number in controlling the theoretical granulation parameters. Recently, \citet{Diaz22} used a subset of the STAGGER grid \citep{Magic2013} to calculate extended timeseries for a number of simulations. In their work, they derive new scaling relations for the standard deviation and autocorrelation time of the timeseries and compare them to those found previously; however they did not investigate the granulation signatures in frequency space. 

A thorough understanding of the underlying granulation background profile in the PDS is invaluable when interpreting the stellar oscillations of a solar-like oscillator (e.g., \citealt{Kallinger2014, Sreenivas24}). \citet{Lundkvist21} investigated the performance of various background model descriptions for a solar simulation, finding that a model with two timescales is optimal. In a sense, the simulations can provide a unique testing ground for such evaluations, where no biases or systematics of observational origin are present -- just the pure granulation signal stemming from the convective motions. In this work, we therefore aim to extend the work of \citet{Lundkvist21} by comparing the performance of several granulation background descriptions to the suite of simulations by \citet{Diaz22}, which are distributed across the Hertzsprung-Russell (HR) diagram. In doing so, we develop a Bayesian nested sampling framework for the granulation background fitting and subsequent model comparison in order to evaluate the merit of various granulation background descriptions extensively used in the scientific community. 

In Sect.~\ref{sec:TheoryData} we present the suite of simulations from \citet{Diaz22} and the theoretical background for describing the granulation background signal. Afterwards, in Sect.~\ref{sec:Methods}, we outline the Bayesian nested sampling methodology used in this paper. Sect.~\ref{sec:Results} contains the resulting model preferences across the suite of simulations and the recovered scaling of the granulation parameters with \numax. To test the developed framework, in Sect.~\ref{sec:Doris} we apply it to the solar analogue KIC 8006161 (also known as Doris) and the Sun, before evaluating the performance of the models. Lastly, further discussions and closing remarks on the future applications are made in Sects.~\ref{sec:Discuss} and \ref{sec:Conclusion}.

\section{Describing the granulation background in 3D simulations}\label{sec:TheoryData}

\begin{figure}[t]
    \resizebox{\hsize}{!}{\includegraphics[width=\linewidth]{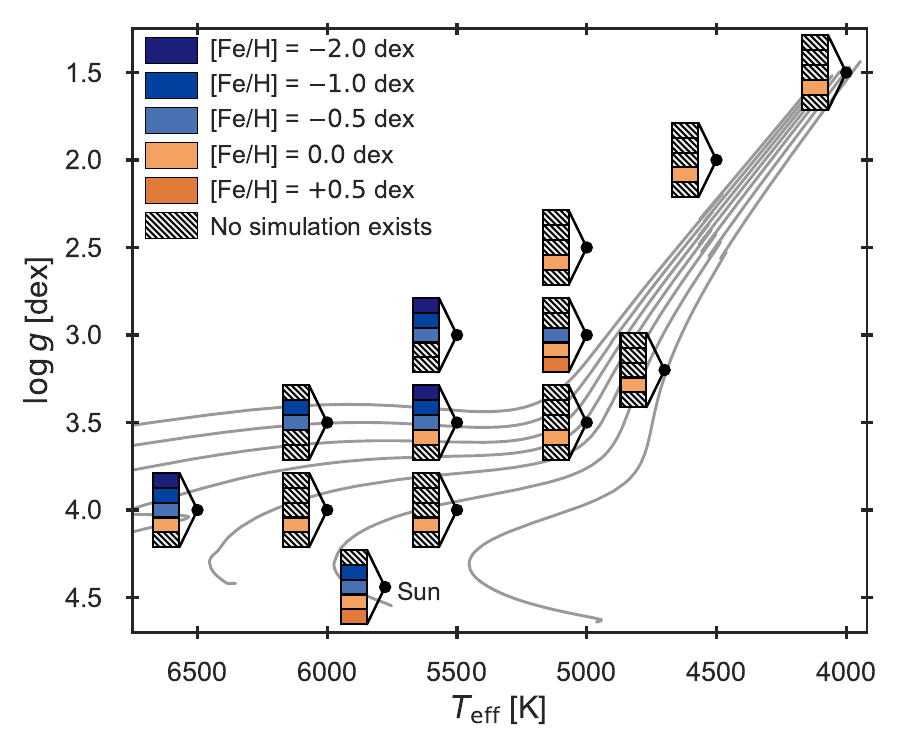}}
    \vspace*{-8mm}
    \centering
    \caption{Kiel diagram displaying the simulation suite from \citet{Diaz22}. The available simulations are plotted for their associated metallicity in vertical stacks. The target (\teff, \logg) positions for each cluster is indicated by the connected black point. Note that the actual temperature of the relaxed simulations differs slightly from the target value. A sample of solar metallicity stellar evolution tracks with masses ranging from 0.8-2.0 $M_\odot$ have been overplotted to guide the eye.}
    \label{fig:Kiel_overview}
    \vspace*{-3.5mm}
\end{figure}
The suite of simulations that we consider originate from the work of \citet{Diaz22}, wherein the detailed creation, descriptions, and studies of the simulations can be found (see e.g. their Table 1). In the present work we only consider the resulting timeseries (Bigot et al. in prep), which are obtained from simulations uniquely suited to our study for two reasons. 

Firstly, they have extremely extended durations, having been run for more than 1000 convective turnover times -- which is defined as the time it takes a convective element to complete an upward (or downward) motion from the bottom to the top (or vice versa) of the simulation box (\citealt{Dethero24}, Eq.~4). Additionally, a simulation snapshot is stored every 60 s for the main-sequence simulations while for the simulations of the most evolved red-giant stars $\approx30$ minutes. This relatively small sampling interval ensures that the Nyquist frequency (the point for which any frequency below is free of aliasing distortions due to the sampling, \citealt{Shannon1949}) of the resulting timeseries is significantly higher than $\numax$.

Secondly, due to the compressible nature of the simulations used in this study, the stochastic excitation of standing sound waves is a natural phenomenon that occurs. The resonance modes of the simulation domain/box are referred to as `box modes' and present as large amplitude Lorentzian peaks in the simulations, which may affect the underlying granulation profile. For the simulation suite of \citet{Diaz22}, the box modes have been artificially damped and removed from the timeseries. The combination of the above makes these simulations particularly apt for providing sufficient resolution as well as avoiding the influence of the prominent box modes in the PDS.

Figure~\ref{fig:Kiel_overview} shows the distribution of the simulations in a Kiel diagram (spectroscopic HR diagram). There are 27 simulations spread out as indicated in the figure between 13 different positions in the Kiel diagram (5 different potential metallicity bins per location). Each simulation provides a timeseries of the horizontally averaged radiative flux as described by \citet{Ludwig06}. We rescale the radiative flux of the simulation box to the entire stellar surface following the approach described by \citet{Trampedach98} and \citet{Ludwig06}, and subsequently calculate the PDS according to \citet{Handberg14}.
\begin{figure}
    \resizebox{\hsize}{!}{\includegraphics[width=\linewidth]{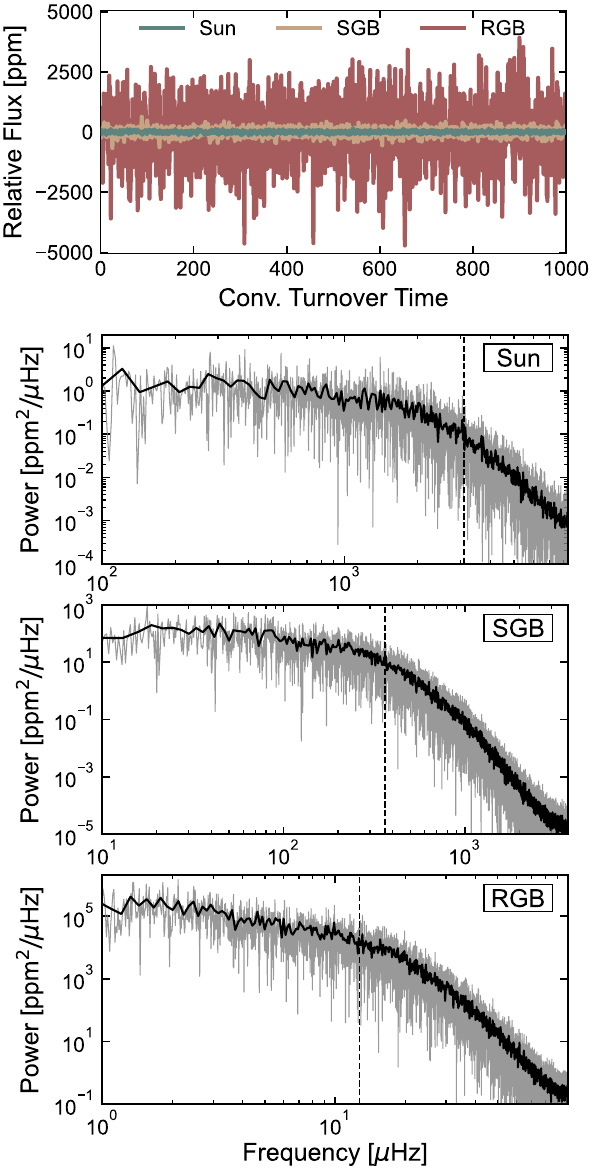}}
    \vspace*{-3mm}
    \centering
    \caption{Timeseries and PDS of three simulations from \citet{Diaz22}, as a reformatted version of their Fig.~3. The timeseries are plotted as a function of convective turnover times and coloured as indicated by the legend. For each PDS, the raw spectrum is shown in grey with a binned version over-plotted in black. The vertical dashed lines indicate the values of $\numax\simeq3094, \  363, \ 12.7$ \muHz for the solar, subgiant (SGB), and red-giant (RGB) simulation, respectively.
    }
    \label{fig:TS_PDS}
    \vspace*{-2mm}
\end{figure}
Figure~\ref{fig:TS_PDS} shows the timeseries and PDS for a solar, subgiant (SGB) and red-giant (RGB) simulation. Contrasting to observed solar-like stars, one can notice the absence of a rise in power at very low frequencies (the `activity') and the lack of an oscillation excess near the indicated \numax. Furthermore, due to the prolonged timeseries that we consider, the granulation profile of the simulations extends to the very high frequency regime with low power.

\subsection{Models for the granulation background signal}\label{subsec:ModelIntro}
Inferring the characteristics of stellar granulation has been done by using various combinations of background components to model the granulation in the PDS. \citet{Harvey85} use a profile where an inverse term with a specific exponent $1/\left(1+(\pi \tau \nu)^{2}\right)$ controlled the decline of the power in the PDS with increasing frequency $\nu$ for a given characteristic timescale $\tau$. \citet{Harvey85} adopted an exponent of 2 (i.e. a Lorentzian function, following from the Fourier transform of the assumed granulation signal in the time domain), however the choice of exponent has a direct consequence on the shape of the background component, which may introduce systematics if the exponent is not properly accounted for. Each component is thereby described by an amplitude and associated characteristic frequency $\nu^l=(1/2\pi\tau)^l$, and an exponent $l$. In the following, we present the models for the granulation background signal that are considered in this work, all summarised in Table~\ref{tab:Models}.

\citet{Kallinger2014} performed an evaluation of various background models for main sequence and red-giant stars using data from the \kepler mission \citep{Borucki10}. They concluded that their model F performed the best across their sample, dubbed a super-Lorentzian function, which has two separate components for the granulation background. Given the methodology and stellar sample studied in \citet{Kallinger2014}, the exponents of this super-Lorentzian function were found to be indistinguishable from 4. Moreover, this value enabled the background model to reproduce the granulation profile in an extended region around the oscillation excesses of the red giants studied. In this work we have, however, instead chosen to use their model H which is a generalised version of model F, where the exponents can vary freely. This avoids limiting the explored parameter space by choosing model F, which exists as a nested model inside the generalised model H (i.e. F is a special case of model H where the exponents equal 4). Additionally, to contrast the performance of single-component vs. multi-component models, the single-component model, denoted D, is also included.

\citet{Lundkvist21} evaluated all the background models of \citet{Kallinger2014} for a 3D hydrodynamical simulation of the Sun from the CO$^5$BOLD code \citep{Freytag12}. In doing so, they also proposed a new hybrid background model with a single amplitude but two characteristic frequencies describing it, one with a locked exponent of 2 and the other one free. In this work we generalise this model by allowing the exponent to vary, denoting it model J. The exponent previously locked came from the expectations of a traditional Harvey profile, yet both exponents are expected to match the predictions from Kolmogorov theory of $l\sim5/3$ and $k\sim17/3$ \citep{Krishan91, Hirzberger97, Krishan02}. These expectations are incorporated through the priors outlined in Sect.~\ref{subsec:Priors}.

Lastly, during the preparations for this work, we found that the PDS of several simulations showed a potential bump in power beyond \numax (seen e.g. for the RGB star in Fig.~\ref{fig:TS_PDS} at $\nu\approx16$~\muHz). This hypothetical third component of the granulation is not clearly seen in real stars, where it would likely be obscured by either an oscillation excess or the level of the white noise. We chose to hypothesise a background model with three individual components, model T, and to evaluate it on equal grounds to the other models across the suite of simulations, despite the reservations for its realistic feasibility. Such a model was also considered by \citet{Zhou21} (see their Sect.~5.1), finding its performance to be comparable to a two-component model such as H.

For each model seen in Table~\ref{tab:Models}, we have an underlying expectation that the components appear at gradually increasing frequency regimes. For example for model H, which is thoroughly discussed in \citet{Kallinger2014}, the expectation is that both components lie below \numax -- the first at lower frequencies and the latter just prior to the oscillation excess. These expectations are encoded in our priors to be presented in Sect.~\ref{subsec:Priors}.

\begin{table*}
\renewcommand{\arraystretch}{2.3}
\centering
\caption{Background models used in this work, listed in order of increasing complexity.}
\begin{tabular}{lcccc}
\hline
\multicolumn{1}{c}{\thead{Model}} & \multicolumn{1}{c}{\thead{Functional form}} & \multicolumn{1}{c}{\thead{Free parameters}} &  \multicolumn{1}{c}{\thead{No. of components}} & \multicolumn{1}{c}{\thead{Reference}} \\
\hline 
D & {\Large $\frac{a^2/b}{1+(\nu/b)^{l}}$} & $a,b,l$ & 1 & \citet{Kallinger2014} \\
J & {\Large $\frac{a^2}{1+(\nu/b)^{l} + (\nu/d)^{k}}$} & $a,b,d,l,k$ & Hybrid & \citet{Lundkvist21} \\
H & {\Large $\frac{a^2/b}{1+(\nu/b)^{l}}+\frac{c^2/d}{1+(\nu/d)^{k}}$} & $a,b,c,d,l,k$ & 2 &\citet{Kallinger2014} \\
T & {\Large $\frac{a^2/b}{1+(\nu/b)^{l}}+\frac{c^2/d}{1+(\nu/d)^{k}} + \frac{e^2/f}{1+(\nu/f)^{m}}$} & $a,b,c,d,e,f,l,k,m$ & 3 &This work \vspace*{1mm}
\\
\hline
\end{tabular}
\vspace*{-1mm}
\tablefoot{The model name, functional form, summarised free parameters, number of components, and reference is given by the table. The parameters $a$, $c$, and $e$ are amplitudes in ppm, while $b$, $d$, and $f$ are the associated characteristic frequencies in \muHz, respectively. The exponents of the characteristic frequencies are denoted $l$, $k$, and $m$.}
\label{tab:Models}
\vspace*{-2mm}
\end{table*}

\subsection{Describing a stellar power-density spectrum}\label{subsec:GranBack}
The aim of this work is to evaluate the performance of background models on simulated models of convection. However, while the simulations provide a unique testing ground, the goal is, in time, to proceed to stellar data. To this end, we wish to compare the model performance while they take the functional form that approximates observational data. This means, that although the simulations do not contain stellar activity, rotation or stellar pulsations, we include additional terms to describe them. When applying the framework to the simulations, we can in turn set priors to reflect the non-existence of these terms (see Sect.~\ref{subsec:Priors}). Here, we introduce the various components and their functional form, before arriving at the complete expression that we fit to the data.

Due to the sampling effects introduced by integrating in discretised measurements of the otherwise continuos stellar variations, a partial cancellation of the signal at higher frequencies occurs \citep{Chaplin11b}. It was found by \citet{Kallinger2014} to influence both the granulation amplitudes and timescales, especially for stars with higher values of \numax. This damping is called apodization and is a function of frequency and the Nyquist frequency $\nu_\mathrm{Nyq}$ of the timeseries, calculated as,
\begin{equation}\label{eq:Apodization}
    \eta(\nu,\nu_\mathrm{Nyq}) = \mathrm{sinc}\left(\frac{\pi}{2}\frac{\nu}{\nu_\mathrm{Nyq}}\right).
\end{equation}
Since we are interested in the integrated granulation amplitudes, the apodization must be accounted for if present. For real stars (such as in Sect.~\ref{sec:Doris}) the apodization factor  $\eta(\nu,\nu_\mathrm{Nyq})$ is calculated following Eq.~\ref{eq:Apodization} and applied as a multiplicative factor to the background model during the sampling. However, in the case of the simulations, the apodization is not applied. The simulation snapshots are saved instantaneously at a particular point in time (i.e. at infinitely sharp times), thus the apodization stemming from the integration time is therefore not present. 

In stellar data, a rise in power is seen in the low frequency regime. This is often attributed to a combination of stellar activity, rotation peaks, and instrumental effects. We chose to simply describe it by a Harvey-profile with a fixed exponent of 2, amplitude $a_2$, and a characteristic frequency $b_2$ \citep{Kallinger2014, Samadi19}:
\begin{equation}\label{eq:Activity}
        \mathcal{M}_\mathrm{act} = \frac{a_2^2/b_2}{1+\left(\frac{\nu}{b_2}\right)^2} \ .
\end{equation}
Solar-like oscillators show an oscillation excess, presenting itself as a forest of individual Lorentzian peaks on top of the granulation background \citep{Chaplin13}. This excess may be assumed Gaussian and centered at the value of \numax \citep{Bedding14}, with $P_\mathrm{osc}$ describing the height and $\sigma$ the standard deviation:
\begin{equation}\label{eq:Oscs}
        \mathcal{M}_\mathrm{osc} = P_\mathrm{osc}\exp{\frac{-(\nu-\numax)^2}{2\sigma^2}} \ .
\end{equation}
Throughout this work, when a calculation of \numax is required we calculated it from the asteroseismic scaling relation \citep{Chaplin13}:
\begin{equation}
    \numax\simeq \frac{g}{g_\odot}\left(\frac{\teff}{T_\mathrm{eff,\odot}}\right)^{-1/2}\nu_\mathrm{max,\odot} \ .
\end{equation}
Lastly, the white noise level of a real star, in the simulations the end of the granulation slope, is described by a constant offset $W$. The constant noise component $W$ is unaffected by the apodization sampling effect, and thus no correction by $\eta(\nu, \nu_\mathrm{Nyq})$ is applied to it. 

Combining all the components outlined above, we arrive at the complete expression that is fit to a given PDS:
\begin{equation}\label{eq:Fullfit} 
    M(\nu) = \eta(\nu,\nu_\mathrm{Nyq})^2 \left[\xi\mathcal{M}_\mathrm{gran} + \mathcal{M}_\mathrm{osc} + \xi_\mathrm{act}\mathcal{M}_\mathrm{act}\right] + W \ .
\end{equation}
Here, a given background model from Table~\ref{tab:Models} is denoted by $\mathcal{M}_\mathrm{gran}$. We reiterate that the apodization factor $\eta(\nu,\nu_\mathrm{Nyq})^2$ defined in Eq.~\ref{eq:Apodization} is not applied when considering the simulations. The factor $\xi$, as described in \citet{Kallinger2014}, normalises the model components such that the square of the amplitude reflects the area under the function in the PDS\footnote{$\xi$ ensures that $\int_0^\infty(\xi/b)/[1+(\nu/b)^n] \ \textup{d}\nu = 1$, satisfying Parseval's theorem such that the area under the function in the PDS equals $a^2$, which in turn equals the variance in the timeseries \citep{Kallinger2014}.}. For a fixed integer value exponent, $\xi$ has analytical solutions; which for an exponent of $2$ results in $\xi=\xi_\mathrm{act}=(2/\pi)^2$, explaining the multiplication factor on the activity term in Eq.~\ref{eq:Fullfit}. When the exponents are freely varied, as is the case for all the background models included in this work, $\xi$ does not have an analytical solution and has to be calculated after the fit has been carried out. An approximation of the $\xi$ factors is given in \citet{Karoff13},
\begin{equation}\label{eq:KaroffNorm}
    \xi \approx 2\alpha\sin\left(\frac{\pi}{\alpha}\right), 
\end{equation}
where $\alpha$ denotes the inferred exponent of a given granulation component (as seen in Table~\ref{tab:Models}). Equation \ref{eq:KaroffNorm} holds for all our background models except model $J$ as it is the hybrid model. In this case, we may instead 
numerically integrate the obtained fit solutions to evaluate the normalisation factor $\xi$ and obtain the normalised model amplitude.

\section{Bayesian nested-sampling}\label{sec:Methods}
We sample the background models in a Bayesian manner. Bayes theorem states that the posterior probability density for a set of parameters $\theta$, given the observed data $D$ and model $M$ is defined as, 
\begin{equation}\label{eq:Bayes}
    p(\theta|D,M) = \frac{p(\theta | M)p(D|\theta,M)}{p(D|M)}.
\end{equation}
Here, $p(\theta|M)$ describes the priors on the model parameters to be introduced in Sect.~\ref{subsec:Priors} and $p(D|\theta,M)$ is the likelihood (see Sect.~\ref{subsec:likelihood}). The factor $p(D|M)$ normalises the numerator, and is known as the marginal likelihood or evidence $\mathcal{Z}$ for the data $D$ given model $M$. It is obtained through an integration over all model parameters, 
\begin{equation}\label{eq:Evidence}
    p(D|M) = \mathcal{Z}= \int_\theta p(\theta|M)p(D|\theta,M) \ \textup{d}\theta \ .
\end{equation}

The approach for performing the sampling is to build a framework using nested sampling \citep{Skilling04}. The benefits of nested sampling is that it is able to handle multi-modal posteriors (see e.g. \citealt{Cai21};\citealt{Dittmann24}), ensuring convergence to global as opposed to local solutions. Furthermore, it can provide an estimate of $\mathcal{Z}$ by approximating the integral of the prior volume, meaning we obtain estimates of both the evidence and samples of the posterior simultaneously. This is contrary to e.g. Markov-Chain Monte Carlo methods which do not generally produce the model evidence and only return samples that are proportional to the posterior. The developed framework uses the software \texttt{Dynesty}\footnote{The background of nested sampling in \texttt{Dynesty} and the documentation can be found here: \url{https://dynesty.readthedocs.io/en/stable/overview.html}.}, which is implemented in Python and developed by \citet{Speagle20}. As nested sampling provides an estimate of the evidence, the change in evidence between iterations may be used to set well-defined convergence criteria for the sampling; set to $\Delta\log(\mathcal{Z})=0.1$ for all purposes of this work, reflecting a strict convergence criteria resulting in increased precision for the obtained evidences.

\subsection{Likelihood function}\label{subsec:likelihood}
The likelihood function describing independent frequency data in the power spectrum is combined with the $\chi^2$ probability distribution function with $2$ degrees of freedom  \citep{Anderson1990,Appourchaux03, Handberg11},
\begin{align}
    \ln\mathcal{L} &= \ln p(D|\theta,M) = \sum_i \ln\left(f(D_i,\theta,M_i) \right), \label{eq:loglike}\\
    f(D_i,\theta,M_i) &= \frac{1}{M_i(\theta)}\exp\left(-\frac{D_i}{M_i(\theta)}\right) \label{eq:chi2}.
\end{align}
Here, $D$ denotes the data, $\theta$ the model parameters and $M$ the given model described by Eq.~\ref{eq:Fullfit}. The index $i$ refers to individual frequency bins in the data, with $D_i$ and $M_i(\theta)$ representing the observed/simulated and modelled (power predicted by a given granulation background) at the $i$-th frequency bin, respectively. Equations~\ref{eq:loglike} and \ref{eq:chi2} are combined to provide the log-likelihood expression to be used in this work.

\subsection{Mixed-model likelihood approach}\label{subsec:mixedmodlik}
\begin{figure}[t]
    \resizebox{\hsize}{!}{\includegraphics[width=\linewidth]{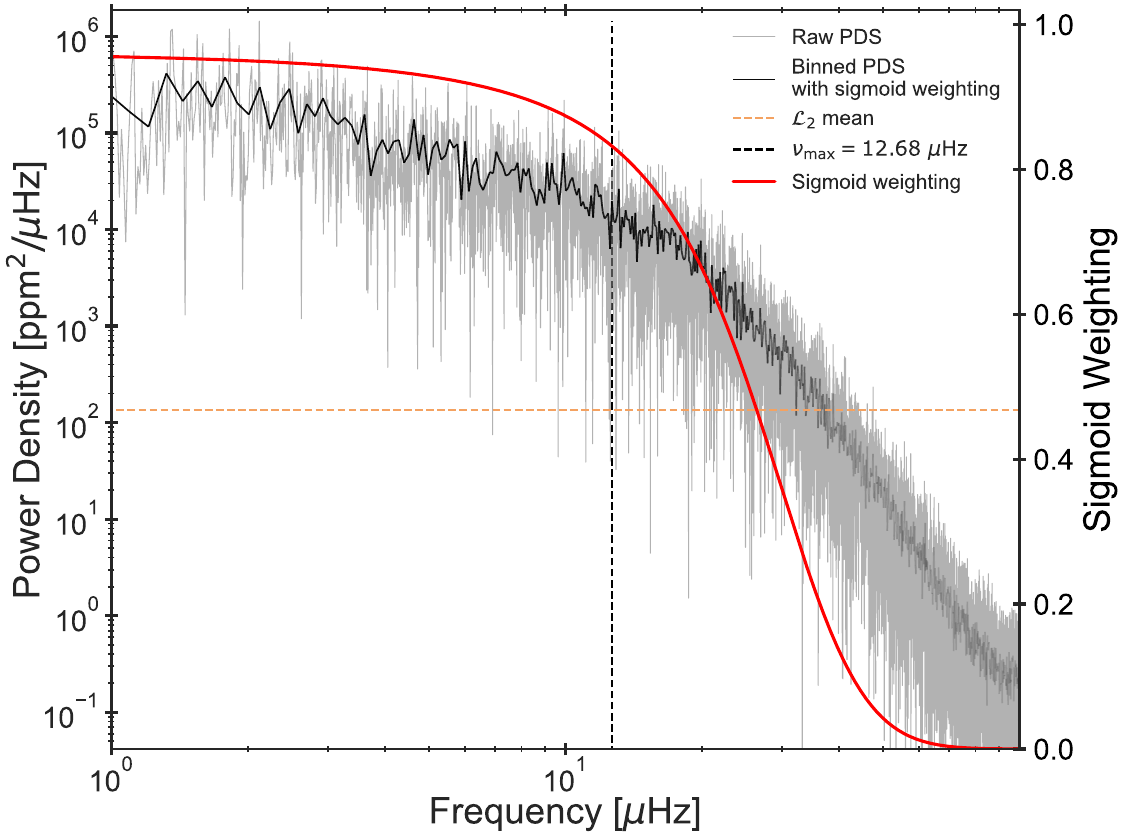}}
    \vspace*{-6mm}
    \centering
    \caption{PDS with sigmoid weighting for the mixed-model likelihood approach for a red-giant simulation. The raw spectrum is shown in grey with a binned version overplotted in black. The opacity of the binned PDS is set by the value of the sigmoid weight, indicated by the red profile gradually decreasing from 1 towards 0 as indicated by the right-hand axis. The mean of the constant $\chi^2$ likelihood $\mathcal{L}_2$ is indicated by the orange horizontal dashed line. The \numax of the simulation is indicated by the black vertical dashed line.}
    \label{fig:PDS_Sigmoid}
    \vspace*{-2mm}
\end{figure}
In Fig.~\ref{fig:TS_PDS}, components in the PDS at very high frequency and low power can be seen. For the evaluation of granulation model performance we wish to focus on the regions of the PDS where the granulation signal in real stars is visible, while being insensitive to the those regions where simulation artefacts could dominate -- such as numerical effects or Lorentzian tails at very high frequencies. For this purpose we implemented a mixed-model likelihood of the form 
\begin{equation}\label{eq:mixedmod}
    \ln\mathcal{L_\mathrm{tot}} = p(\nu)\ln\mathcal{L}_1 + (1-p(\nu))\ln\mathcal{L}_2 .  
\end{equation}
Here, $\mathcal{L}_1$ is the likelihood including a given background model from Eq.~\ref{eq:loglike}. In the second term, $\mathcal{L}_2$ is assumed to be a standard $\chi^2$ distribution with a constant mean set to 1$\%$ of the power near \numax. The factor $p(\nu)$ is a sigmoid function, with upper and lower asymptotes of 1 and 0, respectively. The plateau and slope of the sigmoid is set according to \numax of the given simulation. 

Using this sigmoid as the mixture coefficient is a subjective choice acting as an informed prior, whose use in Bayesian analysis is motivated by \citet{Gelman20}. The choice of the sigmoid function ensures that we do not attempt to implement some complex model for the data, but rather allows the use of Harvey profiles to model it, despite them not reflecting all nuances present in the simulations. Using the sigmoid as the mixture coefficient means that at lower frequency the likelihood described by Eq.~\ref{eq:loglike}, which contains our granulation background model, dominates entirely. As we transition past the value of \numax, the sigmoid weight decreases towards 0, allowing $\mathcal{L}_2$ (i.e. the constant $\chi^2$ likelihood) increasing importance. This effectively downweights the contributions of the high frequency components by allowing $\mathcal{L}_2$ to absorb the features of the PDS for which we do not have a good model. Hence, our later inferences on model preference are based on granulation signatures similar to our expectation for real stars and not affected by simulation-specific nuances.

\begin{figure*}[t]
    \includegraphics[width=17cm]{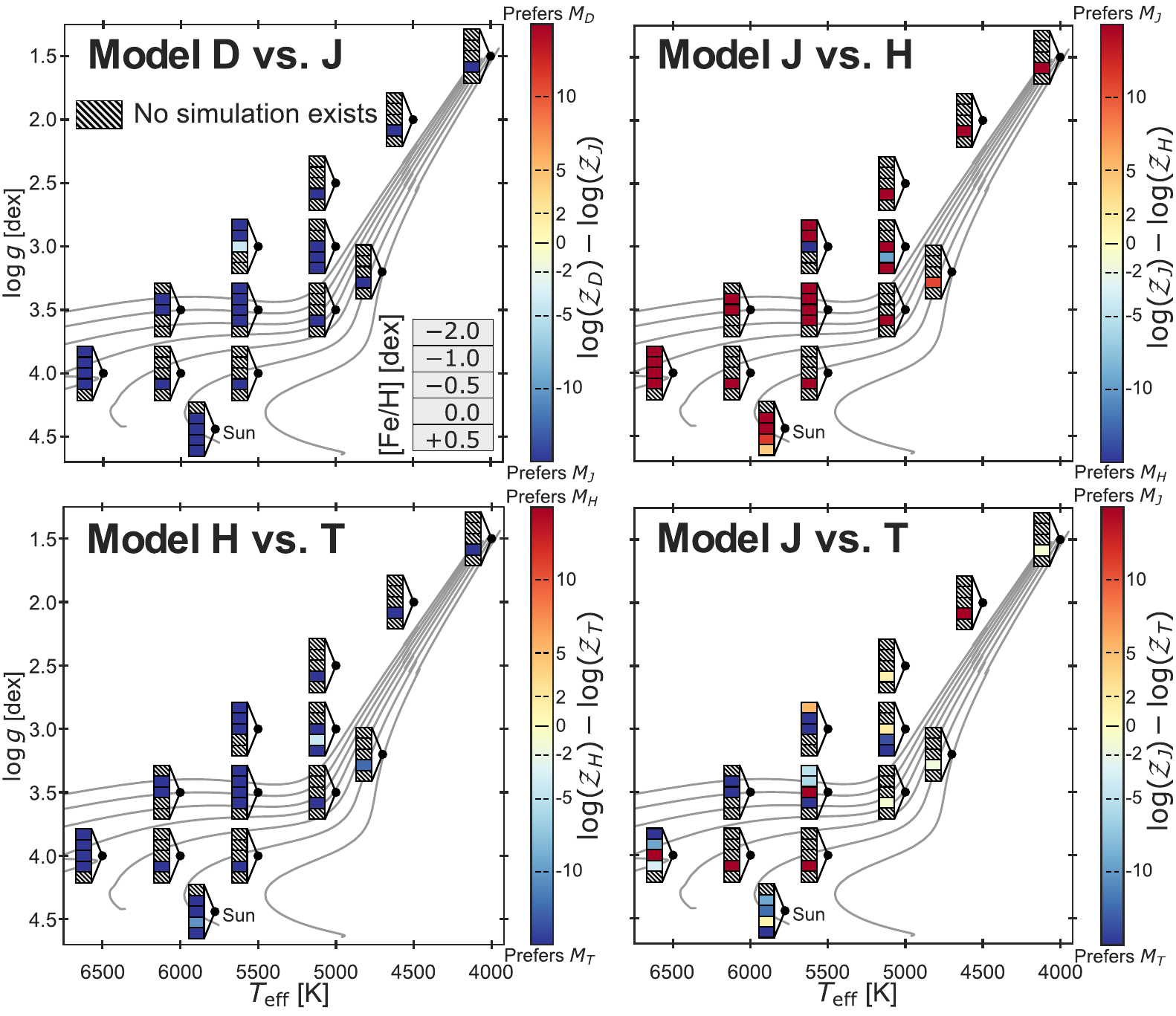}
    \vspace*{-1mm}
    \centering
    \caption{Evidence ratios for the specified model comparisons in increasing complexity. The format of the vertical stacks follow the metallicity convention introduced in Fig.~\ref{fig:Kiel_overview}, but are now coloured according to the evidence ratios as indicated by the colourbar. The Jeffreys' scale for model comparison \citep{Jeffreys61} is indicated on the colourbar, where beyond a $|\textup{d}\log(\mathcal{Z})|\gtrsim 10$ we have overwhelming preference for one model over the other. In the contrary, when $|\textup{d}\log(\mathcal{Z})|\lesssim 2$, no significant preference for either model compared to the other is found. The levels where $|\textup{d}\log(\mathcal{Z})|\simeq 2$ and $5$ indicate significant and decisive preference, respectively.}
    \label{fig:EvidenceRatios}
    \vspace*{-2mm}
\end{figure*}
Figure~\ref{fig:PDS_Sigmoid} visualises the mixed-model likelihood setup by showing the PDS of the RGB simulation previously shown in Fig.~\ref{fig:TS_PDS}. The sigmoid weight, as described above, is dynamically set according to the computed \numax of the simulation. This is indicated in the figure and one may see how the influence of the high frequency components begin to be downweighted after passing the value of \numax.

\subsection{Model Comparison}\label{subsec:ModComp}
A key aspect of this work is being able to compare the performance of the background models across the simulation suite. Making this comparison means asking a different question than done previously. For the fitting of the background models, we wanted to know the optimal parameters $\theta$ given the data $D$ and model $M$. Instead, we now want to determine the probability of model $M_1$ given the data $D$, and compare it to the probability of another model, $M_2$, given the same data. Hence, we write up the posterior odds ratio (see Appendix~\ref{app:A} for a complete derivation), 
\begin{equation}\label{eq:PosteriorOdds}
    \frac{p(M_1|D)}{p(M_2|D)}= \frac{p(D|M_1)}{p(D|M_2)}\frac{p(M_1)}{p(M_2)} = \frac{\mathcal{Z}_1}{\mathcal{Z}_2}\frac{p(M_1)}{p(M_2)} \ .
\end{equation}
The ratio $\frac{p(M_1)}{p(M_2)}$ is the prior odds ratio and reflects our subjective beliefs regarding the considered models. For example, one may consider the single-component model (D) too simplistic to describe the multiple types of variability in stellar data \citep{Kallinger2014,Zhou21,Lundkvist21}, and we would therefore have a prior preference for the two-component model (H) in comparison. Model comparisons are often simplified by considering only evidence ratios, implicitly setting the prior odds $p(M_1)/p(M_2) = 1$ (e.g. \citealt{Handberg11, Muller21}). This choice assumes no prior preference between models, irrespective of whether one is more physically motivated or sophisticated than another. In this work, we adopt the same convention for simplicity, but explicitly acknowledge the underlying assumptions this entails.

\begin{figure*}[t]
    \includegraphics[width=17cm]{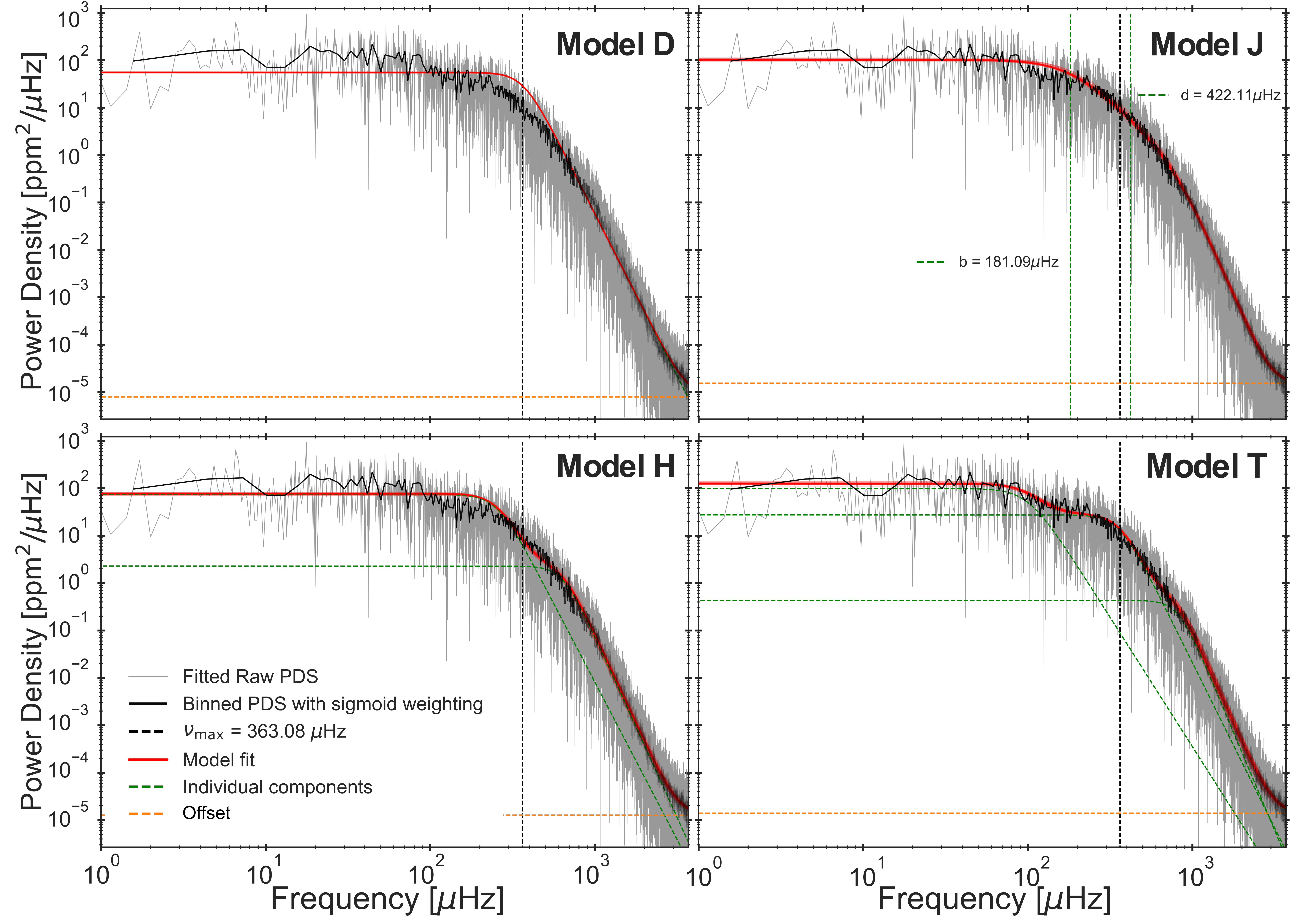}
    \centering
    \caption{Fits of models D, J, H, and T for a subgiant simulation with $\teff=5500$ K, $\log{g}=3.5$ dex and $\FeH=0.0$ dex. The models are fitted to the unbinned PDS shown in grey, but for clarity, a binned version is overplotted indicating also the sigmoid weighting as in Fig.~\ref{fig:PDS_Sigmoid}. The model is plotted in red using the median of the obtained posteriors for each fit parameter. Additionally, 50 randomly drawn samples from the posteriors are used to repeatedly plot the model alongside, as an indicator of the scatter. For models H and T, the individual components are plotted as green dashed profiles. For model J, instead the values of the two characteristic frequencies are shown as green vertical dashed lines. The value of \numax is indicated by the black vertical dashed line and the offset (as the power of the last points of the simulation PDS) by the orange horizontal dashed line.}
    \label{fig:PDS_fits}
    \vspace*{-2mm}
\end{figure*}

\subsection{Priors}\label{subsec:Priors}
In Bayesian analysis, the prior is used to encode our subjective pre-existing belief on the given parameter. The priors that we use are not restrictive, i.e. we allow the prior volume to be large despite the associated computational costs. This choice is made in light of the current limited understanding of the granulation background and the parameters describing its component(s), particularly in the context of the simulations. 

The priors for the amplitudes and timescales are set as log-normal distributions, limiting the parameters to be positive, but allowing for an extended tail of the distribution to higher values. \citet{Kallinger2014} derived various scaling relations for the granulation amplitudes and characteristic frequencies in their work, which we generally rely on to inform our priors, with certain nuances depending on the model (see Appendix~\ref{app:B}). Normal distributions are assumed for the exponents, centered on the optimal values found by \citet{Kallinger2014} and \citet{Lundkvist21}, again depending on the model. In Sect.~\ref{subsec:GranBack} the inclusion of activity and oscillation terms was discussed. We set log-normal priors that reflect their non-existence in the simulations, i.e. an amplitude $\sim$ 0. For an overview of the specific priors for each model and their parameters, see Appendix~\ref{app:B}.

\section{Model preference across the simulation suite}\label{sec:Results}
We applied the framework outlined throughout Sects.~\ref{sec:TheoryData} and \ref{sec:Methods} to all simulations seen in Fig.~\ref{fig:Kiel_overview}. We thereby obtain the posterior distributions for the model parameters, adopting the best-fit estimate as the median of the given distribution with uncertainties as the 16th and 84th percentiles. Furthermore, as outlined in Sect.~\ref{subsec:ModComp}, the nested sampling provides the model evidence $\mathcal{Z}$ when sampling a given model from Table~\ref{tab:Models}. This allows us to evaluate the model preferences across the simulation suite.

Figure~\ref{fig:EvidenceRatios} shows the evidence ratios (i.e. assuming the prior odds ratio $p(M_1)/p(M_2)=1$) of all simulations for the indicated comparisons. One may immediately notice that model D is not preferred over J for any simulation. Model J is preferred over model H in the vast majority of cases, with a few exceptions. Comparing model H to T, we see that model T is clearly preferred except for a few simulations where they perform similarly. A three-component model is thus generally preferred over a two-component model for the simulations, despite down-weighting the importance of the high frequency components. Lastly, we compare the hybrid model J to T. Here the situation is complicated, where for some simulations model J is overwhelmingly preferred, while for others it is the opposite. A notable number of simulations ($\sim$8-10) show no significant preference for either model. We note that the apparent non-monotonic model preference in some panels for the solar-like simulation cluster likely arises from the shorter timeseries duration of the $\FeH = 0.0$ simulation ($\sim$5 days) compared to the others ($\sim$9–10 days). Lastly, although the Bayesian evidence marginalises over the number of free parameters (see Eq.~\ref{eq:Evidence}), it is interesting to see how model J outperforms H and often T while containing fewer free parameters.

\begin{figure*}[t]
    \includegraphics[width=17cm]{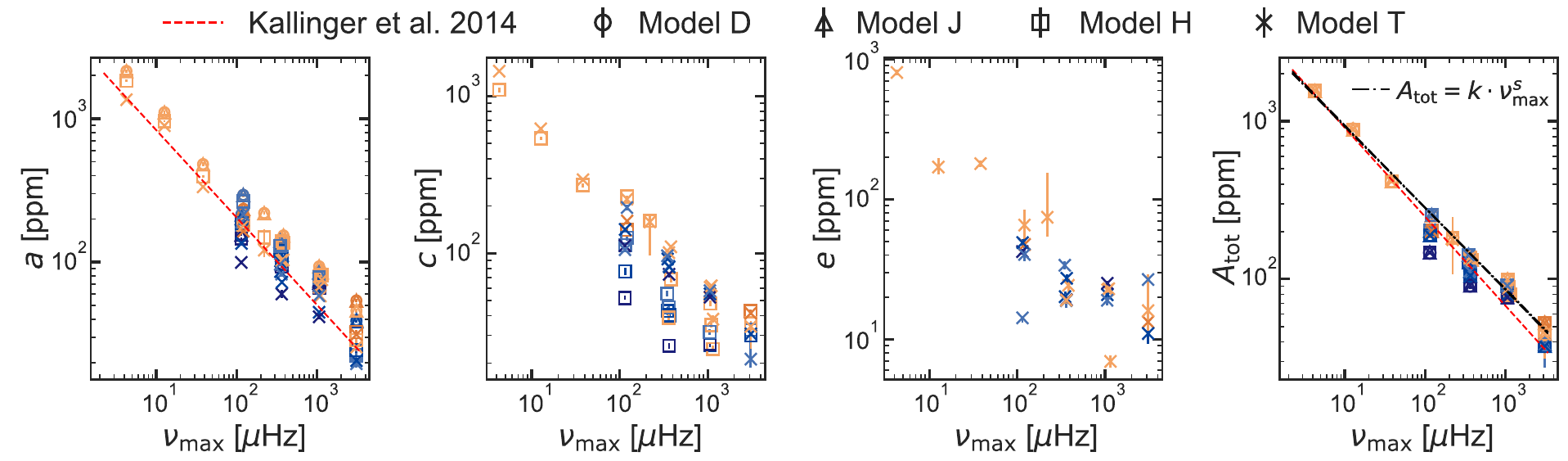}
    \vspace*{-1mm}
    \centering
    \caption{Granulation amplitudes, normalised by $\xi_i$ as explained in Sect.~\ref{subsec:GranBack}, obtained as the median of the posterior with uncertainties as the 16th and 84th percentiles (often too small to be seen). The colours indicate metallicity following the convention in Fig.~\ref{fig:Kiel_overview}. The different symbols indicate the various background models. The panels show the amplitudes $a$, $c$, and $e$, plotted for the associated models. The last panel shows the total granulation amplitude, $\mathrm{A}_\mathrm{tot}$. In the panels for which direct comparisons to the relations derived in \citet{Kallinger2014} can be made, they are plotted as red dashed profiles. Three power-law fits are shown as the black dot-dashed profiles for the total granulation amplitudes, which completely overlap.}
    \label{fig:GranAmps}
    \vspace*{-2mm}
\end{figure*}

\begin{figure*}
    \includegraphics[width=17cm]{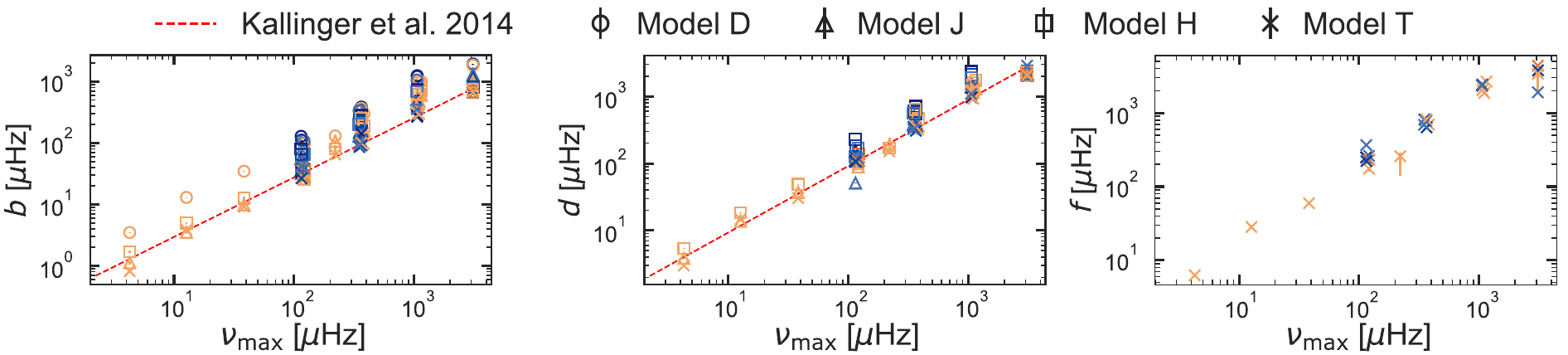}
    \vspace*{-1mm}
    \centering
    \caption{Characteristic frequencies obtained as the median of the posterior with uncertainties as the 16th and 84th percentiles (often too small to be seen). The colours indicate metallicity, following the convention in Fig.~\ref{fig:Kiel_overview}. The different symbols indicate the various background models. In the panels for which direct comparisons to the relations derived in \citet{Kallinger2014} can be made, they are plotted as red dashed profiles.}
    \label{fig:GranFreqs}
    \vspace*{-2mm}
\end{figure*}
Figure~\ref{fig:PDS_fits} shows the samples of models D, J, H, and T for a subgiant simulation as an example, for which the model evidence ratios of Fig.~\ref{fig:EvidenceRatios} show a clear preference for model T. The evidence provides statistically rigorous conclusions regarding model preference, and a visual inspection reveals that the single-component model D fails to represent the data, as it reproduces neither the amplitude level nor the correct shape of the decline in power. Similarly, model H struggles to accurately represent the granulation signal with its two individual components; placing one term below \numax and one above. Model J shifts its upper characteristic frequency past \numax and thereby better represents the data. However, superior to all others in this case is model T, which places two of its components below \numax and the last above. In cases such as shown in Fig.~\ref{fig:PDS_fits}, model T clearly fits the data better than the remaining models, despite the contributions from the high frequency components being down-weighted by the mixed-model likelihood. This interpretation is confirmed by the model evidences we obtained for each sampled background model for this simulation. However, for a number of simulations the third component of model T produces an unnecessarily complicated fit with more free parameters -- which is penalised in the marginalisation over all parameters in Eq.~\ref{eq:Evidence} -- causing models J or H to be preferred as presented in Fig.~\ref{fig:EvidenceRatios}. 

Given the results presented above, we note that models J, H, and T all have merit for specific simulations. In the following, we evaluate the behaviour of the granulation parameters across the simulation suite for the models, before continuing to the application of the framework to real stellar data in Sect.~\ref{sec:Doris}.

\subsection{Granulation amplitudes and characteristic frequencies}
An essential output of the nested sampling are samples from the posterior for each fitted parameter. This allows us to inspect the behaviour of the various parameters in the components of the granulation background profiles across the suite of simulations. In Figs.~\ref{fig:GranAmps} and \ref{fig:GranFreqs}, we show the median of the posterior with uncertainties as the 16th and 84th percentiles for the amplitudes and characteristic frequencies, respectively. For all four background models the amplitudes decrease with \numax as expected \citep{Kjeldsen11,Mathur11,Samadi13b,Kallinger2014,Diaz22} and generally follow the observationally determined scaling from \citet{Kallinger2014}. Notably, the tentative third component of model T shows a similar dependence on \numax, however with significant scatter.

The total bolometric intensity fluctuation stemming from granulation can be estimated as $A_\mathrm{tot}^2 = C_\mathrm{bol}^2\left(\sum a_i^{2}\right)$, where $C_\mathrm{bol}=(\teff/5934 \ \textup{K})^{0.8}$ and $a_i$ are the normalised amplitude(s) of the given model \citep{Michel09,Ballot11}. Following this, we propagated the errors on the individual normalised amplitudes onto the total granulation amplitude, providing fractional errors in the range $2-10\%$. For model T specifically, we inspected the residuals in the total granulation amplitude. We found that removing the contribution to $A_\mathrm{tot}$ of this tentative third component did not increase the scatter in the residuals, as the amplitude of the third component is order of magnitudes lower than the first and second components. 

We fit a power law of the form $f=k\cdot\numax^s$ to the total granulation amplitudes $A_\mathrm{tot}$ for the solar metallicity $\FeH=0.0$ dex simulations using the \emph{iminuit} package\footnote{\url{https://scikit-hep.org/iminuit/index.html}} with the MINUIT algorithm \citep{MinuitAlg}. Per the results in Fig.~\ref{fig:EvidenceRatios}, we omitted the amplitudes estimated by model D, resulting in three fits for the remaining models with 11 data points each. The results are shown in Fig.~\ref{fig:GranAmps} and found $k=3040\pm 90, \ 3140\pm70, \ 3050\pm100$ and $s=-0.516\pm0.005, \ -0.522\pm0.004, \ -0.515\pm0.006$, for models J, H, and T, respectively. These results show a significant difference to the relation $A_\mathrm{gran}=3335\pm9 \cdot \numax^{-0.564\pm 0.002}$ ppm found by \citet{Kallinger2014}. This is not entirely unsurprising as the underlying methodology and the nature of the data (real stars vs. simulations) are different. Yet, it emphasises how the determined granulation amplitudes depend on the specific dataset and the chosen model to describe the overall granulation background. Lastly, it is clear from the amplitudes seen in Fig.~\ref{fig:GranAmps} that a change in the metallicity affects the obtained granulation amplitudes, generally hinting that a decrease in metallicity reduces the granulation amplitude. This trend was also found by \citet{Diaz22} when inspecting the standard deviation of the simulation timeseries, and seconds the observational findings of \citet{Corsaro17} and \citet{Yu18}.

The characteristic frequencies in Fig.~\ref{fig:GranFreqs} similarly scale with \numax, where $b$ and $d$ follow the relation derived by \citet{Kallinger2014}. The correlation is also seen for the third granulation component of model T. This is interesting and suggests the presence of a genuine granulation component beyond \numax. However, we caution that while the granulation is expected to scale with \numax, so are the simulation box modes and their harmonics \citep{Nordlund01, Zhou21}. While the principle box modes are damped in the simulation suite of \citet{Diaz22}, any tentative residuals of the harmonics could be what the third component of model T seeks to model.

\subsection{Exponents of background models}\label{subsubsec:Exponents}
\begin{table}
\renewcommand{\arraystretch}{1.5}
\centering
\caption{The median values of the exponents recovered from the fitting across the simulation suite.  
}
\begin{tabular}{ccc}
\hline
\multicolumn{1}{l}{\thead{Model}} & \multicolumn{1}{c}{\thead{Median of exponent}} & \multicolumn{1}{c}{\thead{Expectation}} \\
\hline 
D & \thead{$l\simeq5.83$} & \thead{$l\simeq2$} \\
J & \thead{$l\simeq2.4$ \\ $k\simeq6.9$} & \thead{$l\simeq5/3$ \\ $k\simeq17/3$} \\
H & \thead{$l\simeq5.4$ \\ $k\simeq6.7$} & \thead{$l\simeq4$ \\ $k\simeq4$} \\
T & \thead{$l\simeq5.0$ \\ $k\simeq6.1$ \\ $m\simeq7.4$} & \thead{$l\simeq4$ \\ $k\simeq4$ \\ $m\simeq$ unknown} \\
\hline
\end{tabular}
\tablefoot{The expectations for models D and H stem from \citet{Kallinger2014}, and for model J from \citet{Lundkvist21} (see details in Sect.~\ref{subsec:ModelIntro}). For model T, the expected values are not truly expectations based on any literature, but simply a naive extrapolation from the two-component model H.}
\label{tab:Exponents}
\vspace*{-2mm}
\end{table}

The exponents of the characteristic frequencies controls how rapidly the power of the respective granulation component declines. \citet{Kallinger2014} found that for their best-fitting model with two individual components, both exponents were indistinguishable from 4 given the data. \citet{Lundkvist21} discussed the free exponent of their hybrid model for a solar simulation, finding a value of around 6 was preferred, but also noting that for the two-component model H values different from 4 were found. In the following, we briefly summarise the median values obtained in this work.

Table~\ref{tab:Exponents} shows the median value of the exponent across the simulation suite for the components of the given model. Model D, akin to the standard Harvey profile \citep{Harvey85}, but with a free exponent, finds an exponent far higher than the originally proposed value of 2. For model J, both exponents are also recovered at values somewhat higher than the expected from the work of \citet{Lundkvist21}. Model H, which according to \citet{Kallinger2014} should display exponents close to 4, returns exponents for both components that are significantly larger than expected. For the three-component model T, a naive expectation simply extrapolating from the two-component model results in the exponents of the inner two components being higher than expected. However, we stress that the exponents may well change when altering the model description. The third component displays the highest exponent of all, corresponding to the sharpest decline in the power with increasing frequency. Thus, across the simulation suite we find for all tested background model descriptions, that the exponents are higher than expected.

The simulations are a unique testing ground where no oscillation excess or white noise level can affect our investigations. This results in the granulation profile being visible at lower power and higher frequencies than would be feasible to observe for real stars. That we find higher exponents of the granulation components could suggest that for real stars, the oscillations and white noise levels artificially decrease the granulation exponents. Verifying this indication requires further study, including the first application and subsequent performance assessment of models J and T to stellar data.

\begin{figure*}[t]
    \includegraphics[width=17cm]{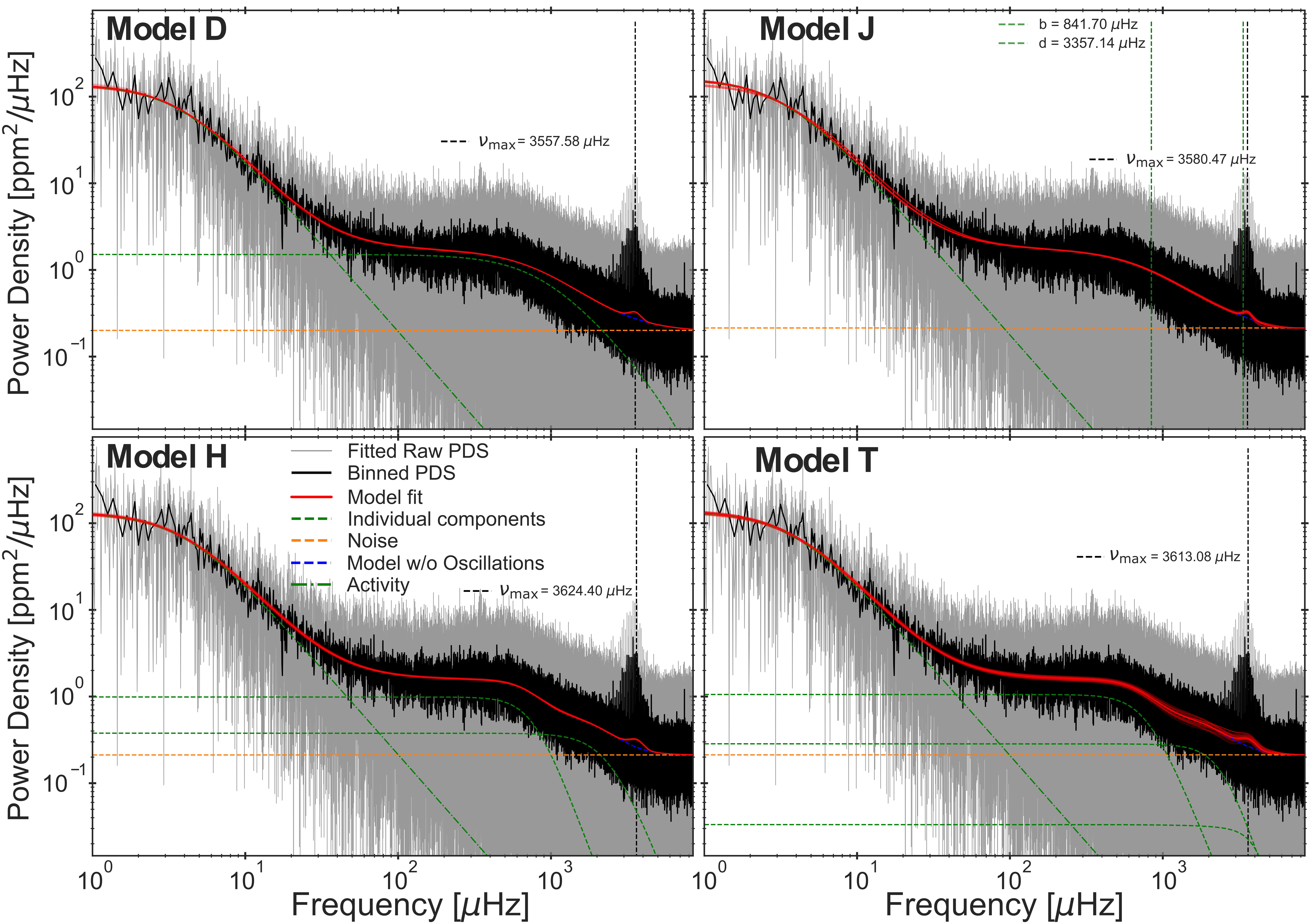}
    \centering
    \caption{Fits of models D, J, H, and T to Doris. The models are fitted to the unbinned PDS shown in grey, but for clarity, a binned version is overplotted. The model is plotted in red using the median of the obtained posteriors for each fit parameter. Additionally, 50 randomly drawn samples from the posteriors are plotted to indicate the scatter. For models D, H, and T, the individual components are plotted as green dashed profiles. For model J, the values of the two characteristic frequencies are shown as green vertical dashed lines. The fitted value of \numax is given and indicated by the black vertical dashed line, while the noise is shown by the orange horizontal dashed line. The activity component is the green dot-dashed line. The fitted model without the influence of the oscillation excess in plotted as the blue dashed profile, visible underneath the oscillation excess.}
    \label{fig:DorisFits}
    \vspace*{-2mm}
\end{figure*}

\section{Extension to stellar data}\label{sec:Doris}
The development of this framework, while focusing on the unique testing grounds provided by the simulations, aims for future application to stellar data. To this end, we test its performance by carrying out the same model comparison as done for the simulations for a solar analogue KIC 8006161, popularly called Doris, and subsequently the Sun itself. The differences here to the methods used for the simulation results in Sect.~\ref{sec:Results}, is that we do not use the mixed-model likelihood to downweigh the high frequency regions described in Sect.~\ref{subsec:mixedmodlik}. Instead, we use the $\chi^2$ likelihood from Eqs.~\ref{eq:loglike} and \ref{eq:chi2}. This choice is motivated by the fact that the granulation background does not extend to very high frequency and low power (as seen in e.g. Fig.~\ref{fig:TS_PDS}); rather, the signal is dominated by an oscillation excess and white noise level. Hence, the priors for the activity and oscillation excess components are changed to reflect their presence (see Appendix~\ref{app:B}), and \numax is now treated as a fit parameter.

\subsection{Doris -- KIC 8006161}\label{subsec:ValDoris}
Doris was observed by \kepler in short cadence in quarters Q5-17 and has $\numax\approx3575$ \muHz, $\teff \approx5488$ K, $\FeH\approx0.34$ dex, and $\log(g)=4.94$ dex \citep{Lund2017}. We used the full short cadence timeseries after applying the KASOC filter \citep{Handberg14}\footnote{Retrieved from the KASOC database: \url{https://kasoc.phys.au.dk}.} and computed the PDS following \citet{Handberg14}. 

Figure~\ref{fig:DorisFits} presents the model fits to Doris. Inspecting the top two panels for models D and J, we see that near the granulation plateau (middle region starting at $\nu\sim100$ \muHz) they perform somewhat similarly. Models H and T are more versatile due to their multiple individual components and, thereby, describe the transition from the activity component (end of initial slope near $\nu\sim30$ \muHz) to the granulation plateau differently. As seen for the simulations, model T places two components below \numax and one just above, featuring a very low amplitude in comparison to the two inner components. Nevertheless, model T uses its third component to lift the granulation background slightly underneath the oscillation excess. By inspecting the obtained values of \numax in Fig.~\ref{fig:DorisFits}, we see clearly that they are affected by the underlying model for the granulation background, which is in agreement with the results by \citet{Sreenivas24}.

\begin{table}
\renewcommand{\arraystretch}{1.5}
\centering
\caption{Evidence ratios for model comparison of Doris (KIC 8006161).}
\begin{tabular}{cc|cc}
\hline
\thead{Model Comparison} & \thead{$\log(\mathcal{Z}_1)-\log(\mathcal{Z}_2)$} & \thead{Model} & \thead{Ranking} \\
\hline 
D vs. J & $-257.61\pm{0.30}$ & D & 4 \\
J vs. H & $-585.21\pm{0.28}$ & J & 3 \\
H vs. T & $-2.94\pm{0.26}$ & H & 2 \\
J vs. T & $-588.15\pm{0.27}$ & T & 1 \\
\hline
\end{tabular}
\tablefoot{The left-hand columns list the pairwise model comparisons and corresponding evidence ratios. For clarity, the right-hand columns indicate the inferred model and its ranking based on overall support from the evidence. The displayed uncertainties have been estimated by propagating the errors of the individual log-evidences.}
\label{tab:DorisZs}
\vspace*{-4mm}
\end{table}

Table~\ref{tab:DorisZs} gives the evidence ratios for the same model comparisons as shown for the simulations in Sect.~\ref{sec:Results}. The picture is quite clear: model J is overwhelmingly preferred to model D, while the same is true for models H or T compared with model J. Models H and T perform similarly, though with a slight preference for model T. Notably, as seen for several simulations in Sect.~\ref{sec:Results}, model T is quantitatively preferred, here for a real star. This puts into question if the preference found for the simulations truly was a real granulation signature beyond \numax and not some box mode residual. Yet, in our work the oscillation excess is somewhat crudely modelled by a Gaussian envelope. We note that it is conceivable that the implementation of a more complex model for the oscillations, like a Voigt profile \citep{Voigt2006} or a mixture model for the likelihood to allow for `noise peaks' \citep{Littenberg15}, could eliminate the preference for model T.

\subsection{The Sun as a star}\label{subsec:ValSun}
Given the results found for Doris, we chose to further test the framework for the Sun. The Sun is very bright, far surpassing what is achievable for other stars, resulting in an extremely low white noise level. In a sense, we may consider the Sun as a star without noise for us to extend our framework to. For this purpose, we used blue-band VIRGO timeseries data \citep{Froehlich88} with a duration identical to that available for Doris ($\simeq 1150$ days), taken during the solar minimum between cycle 23 and 24. The PDS was calculated following \citet{Handberg14}.

\begin{figure}[t]
    \resizebox{\hsize}{!}{\includegraphics[width=\linewidth]{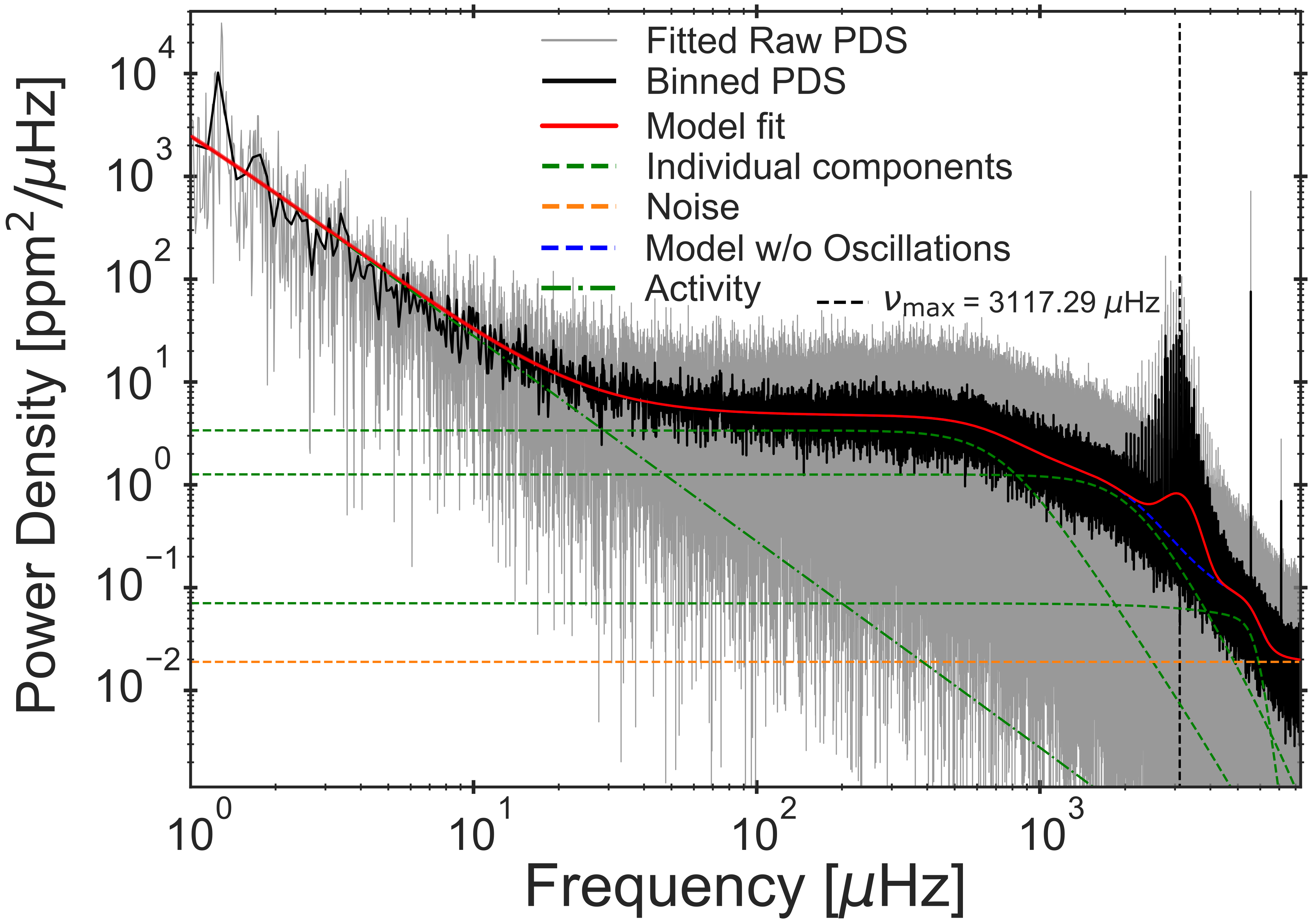}}
    \vspace*{-1mm}
    \centering
    \caption{Fit of model T to $\simeq 1150$ days of VIRGO data of the Sun. The nomenclature of the plot is as explained in Fig.~\ref{fig:DorisFits} and the legend.}
    \label{fig:PDS_SunT}
    \vspace*{-2mm}
\end{figure}
Figure~\ref{fig:PDS_SunT} shows the fit of model T to the Sun, while the presentation of the other fits and model evidences may be found in Appendix~\ref{app:C}. In contrast to Doris in Fig.~\ref{fig:DorisFits}, one can clearly see that the white noise is an order of magnitude lower in this case. In the absence of the white noise, we see that model T finds a clear use for its third component beyond \numax, and the evidence indicates an overwhelming preference for model T. Moreover, allowing a third component enables the two inner components to describe the granulation in the frequency regions according to the scaling relations found by \citet{Kallinger2014}; in Fig.~\ref{fig:SunFits}, it may be seen how model H otherwise shifts its outer component below the oscillation excess in an attempt to better match the presence of power at higher frequencies.

However, while the results found for the Sun corroborate a tentative third granulation component, the nuances mentioned in Sect.~\ref{subsec:ValDoris} persist. For the Sun specifically, high radial order $n$ modes may be observed, which directly affects the shape of the oscillation excess. This could lead to a contribution in lifting the underlying slope at the high frequency end of the oscillation excess, resulting in the preference for model T. An improved model description for the oscillations in the context of investigating the granulation components is thereby further justified for future pursuits.

\section{Discussion}\label{sec:Discuss}
This discussion briefly summarises certain results of the present work and relates them to existing findings in literature, before highlighting potential future directions.

\subsection{Granulation components and faculae}
It was repeatedly seen in Sect.~\ref{sec:Results} that a background model with two characteristic frequencies/components (or more) always outperformed the single-component model. In Fig.~\ref{fig:EvidenceRatios}, this is true for all but a single subgiant simulation. However, for this specific simulation a significant modulation signal is present in the PDS, for which the origin is unknown yet clearly artificial. 

Notably, this preference occurs in simulations with no magnetic activity. \citet{Karoff13} argued that a two-component model is required to describe the signal from stellar faculae, which originate from the presence of magnetic fields. Given that we find that multi-component models are preferred in describing the granulation signal of the simulations, we can infer that stellar faculae is not the only potential source of the secondary component.

\citet{Magic14} performed detailed analysis of the distribution of granule size for Stagger-grid models. They found a broad distribution of granule size that contains two main groups, one smaller than the mean granule size and the other larger (\citealt{Magic14}, Fig.3). Furthermore, when investigating the intensity distribution of the granules, differences in the mean bolometric intensities between these two groups were found, the smaller granules being darker and the larger brighter. As the characteristic timescale and lifetime of the granulation correlates with its size, and variations in the intensities were found, granulation power spectra predicted by such surface convection simulations ought to be multi-component. Further considerations of such findings in 3D hydrodynamical simulations and their synergies with observed granulation signals may provide an avenue for deriving more physically grounded models for the granulation background.

Observationally, \citet{Kallinger2014} found that a single-component model is only applicable if the white noise dominates the background, such that the individual components of the granulation and their shape are hidden. How the model preferences in our framework depend on the signal-to-noise ratio and the resolution of observations will be evaluated when applying the framework to a broader range of real stars (Larsen et al. in prep).

\subsection{Interplay of granulation and oscillation signals}
For future applications, the interplay between the amplitude of granulation and stellar oscillations deserves consideration. \citet{Sreenivas25} studied their wavelength dependence and found that the signal from granulation and oscillations have the same wavelength dependence, avoiding the situation where one would obscure the signal of the other. 

An intriguing region for potential utilisation of the granulation background signal and its connection to the underlying stellar parameters (e.g. $\log(g)$ using the FliPer algorithm by \citet{Bugnet18}), is for stars where the granulation background may be resolved but the oscillations are not. This could potentially happen if significant damping due to rotation or magnetic fields is present for the oscillations \citep{Chaplin11a,Huber11,Campante14}. Furthermore, it could possibly provide an avenue for obtaining a deeper understanding and characterisation of the stars observed by Kepler and TESS \citep{Ricker14} for which no oscillations were or are found. For \textit{Kepler}, \citet{Mathur17} found ${\sim}$2/3 of the observed red giants to be oscillating, and for TESS specifically, \citet{Hey24} highlighted how limitations in data quality and cadence can complicate the detection and interpretation of oscillation signals for subsequent asteroseismic diagnostics. Applying the developed framework to such cases could provide valuable insights into the nuances of the granulation background, without the presence of robust asteroseismic detections.

\section{Conclusion}\label{sec:Conclusion}
This work has extended the coverage of power density spectra from 3D hydrodynamical simulations for which the performance of various granulation background models have been evaluated. Using the suite of 27 simulations from \citet{Diaz22}, a Bayesian nested sampling framework has been developed for model inference and performance comparison for different background model descriptions. This framework is suitable for further application to stellar data beyond the test cases of Doris and the Sun, in the search for accurate granulation background descriptions. The main insights and conclusions of this work are summarised as follows:
\begin{itemize}
    \item Multi-component granulation background descriptions are always preferred to a single-component model. The hybrid model (J), two-component model (H), and three-component model (T) all have merit for future applications in specific regions of the HR diagram. As the simulations have no magnetic activity and a single-component model is never preferred, stellar faculae cannot be the only source of the second background component.
    \item A tentative preference for a third granulation component beyond the value of \numax was found for numerous simulations. Determining whether this signal is a numerical artefact or a potentially real higher frequency granulation component will require further study of stellar data of high quality, and likely also developments on the applied model for the oscillation excess.
    \item KIC 8006161, Doris, was used to test the developed framework on stellar data. Upon comparing the models, no significant difference was found between a two- and three-component model, but the single-component and hybrid models were not favoured in this specific case. The results clearly show that the chosen granulation background description affects the \numax determination significantly, and showed that the developed framework is suitable for future application to observational data. In the unique case of the Sun, overwhelming preference for a three-component model was found, yet it may be due to the assumed model for the oscillation excess or the nuances of solar data. 
\end{itemize}
This work presented the preference between different granulation background models used as literature standards and potential extensions of such. Yet, we caution drawing a direct connection between the conclusions reached here on the basis of simulations and those of real observations. As seen and discussed in Sects.~\ref{sec:Results}, \ref{sec:Doris}, and \ref{sec:Discuss}, the tentative third component of model T may be describing different contributions in the simulations and the observations. For the simulations, it could be reproducing the decrease in the granulation slope beyond \numax, while for the observations it may be handling potential residuals from the oscillation excess. Hence, further rigorous testing and scrutiny of the models showing merit (J, H and T) during application to a wide range of stars is needed to assess the most suitable background description for application to real stars (Larsen et al. in prep).

The future potential of drawing on the information embedded in the granulation signal is manifold. With the continuous observations by TESS and the upcoming PLAnetary Transits and Oscillations of stars mission (PLATO; \citealt{PLATO24}), the wealth of timeseries data for a wide variety of stars is increasing steadily. Developing our understanding of the surface manifestation of convection may unveil independent ways of characterising solar-like stars from such observations, potentially yielding not only inferences on the surface gravity but direct information on stellar radii.

\begin{acknowledgements}
    The authors thank the anonymous referee for the constructive comments made on our paper. JRL wishes to thank the members of SAC in Aarhus and the Sun, Stars and Exoplanets group in Birmingham for comments and discussions regarding the paper. This work was supported by a research grant (42101) from VILLUM FONDEN. MSL acknowledges support from The Independent Research Fund Denmark's Inge Lehmann  program (grant  agreement  no.:  1131-00014B). Funding for the Stellar Astrophysics Centre was provided by The Danish National Research Foundation (grant agreement no.: DNRF106). The numerical results presented in this work were partly obtained at the Centre for Scientific Computing, Aarhus \url{https://phys.au.dk/forskning/faciliteter/cscaa/}. This paper received funding from the European Research Council (ERC) under the European Union’s Horizon 2020 research and innovation programme (CartographY GA. 804752). This paper includes data collected by the Kepler mission and obtained from the MAST data archive at the Space Telescope Science Institute (STScI). Funding for the Kepler mission was provided by the NASA Science Mission Directorate.
\end{acknowledgements}

\section*{Data availability}
The fitting framework is accessible on GitHub and the data available following reasonable request to the first author.

\bibliographystyle{aa.bst} 
\bibliography{bibliography.bib} 

\newpage
\begin{appendix}
\section{Posterior Odds Ratio}\label{app:A}
Model comparison within the Bayesian school of thought means evaluating the posterior probability odds ratio, also referred to as the Bayes factor \citep{Morey16}. To compare model $A$ and $B$ given data $D$, we must evaluate $\frac{p(A|D)}{p(B|D)}$. We may use Bayes theorem to rewrite this as,
\begin{equation}\label{eq:PosteriorOddsApp}
    \frac{p(A|D)}{p(B|D)} = \frac{p(D|A)}{p(D|B)}\frac{p(A)}{p(B)}\frac{p(D)}{p(D)}. 
\end{equation}
The last factor describing the prior of the data thus cancels. The second term is the prior odds ratio and reflects the prior beliefs before considering the data. The first term depends on the data and has to be calculated. For e.g. the numerator this may be considered as a marginalisation over the free parameters $\theta_A$ of a given model $M_A$, 
\begin{align}
    p(D|M_A) &= \int_{\theta_A} p(D,\theta_A|M_A) \ \textup{d}\theta_A \\
    &= \int_{\theta_A} p(D|\theta_A,M_A)p(\theta_A|M_A) \ \textup{d}\theta_A .
\end{align}
Here, we have used the product rule of probability $p(x,y|z)=p(x|y,z)p(y|z)$ to rewrite the integrand. The integrand is now split into two parts, representing the likelihood and the prior, respectively. We may recognise this integration as the marginalised likelihood, i.e. model evidence $\mathcal{Z}$. These steps can be repeated for model $M_B$ with parameters $\theta_B$ and Eq.~\ref{eq:PosteriorOddsApp} may then be written as,
\begin{equation}
    \frac{p(A|D)}{p(B|D)} = \frac{\mathcal{Z}_A}{\mathcal{Z}_B}\frac{p(A)}{p(B)}.
\end{equation}

\section{Overview of priors}\label{app:B}
This appendix specifies the various priors used in this work, compiled in Table~\ref{tab:PriorComp}. They largely rely on the various scaling relations for the granulation amplitudes and characteristic frequencies derived in \citet{Kallinger2014}. For modified and new background models, the expected behaviour of the component is encoded in the prior, but still based on the scaling relations of \citet{Kallinger2014}. E.g. for the third high frequency component of model T, the priors are set to reflect a lower amplitude and higher frequency than the second component. As model J takes a different shape than the rest, we derived our own scaling relation for its amplitude. This was done by first running test-fits with wide priors set by the scaling relations of \citet{Kallinger2014}, before deriving a scaling relation for the amplitude based on the preliminary outcomes, and subsequently using it when producing the results of this work. The exponents of the various models are set following the best-fitting solutions found by \citet{Kallinger2014} and \citet{Lundkvist21}. 
\newpage
\begin{table}
\renewcommand{\arraystretch}{1.5}
\centering
\caption{Log-space parameters for the power law scaling relations from Table~2 of \citet{Kallinger2014}.}
\begin{tabular}{lccc}
\hline
\multicolumn{1}{l}{\thead{Component}} & \multicolumn{1}{c}{\thead{Value of constants}} & \multicolumn{1}{c}{\thead{Value of exponents}} \\ \hline 
1st component & \thead{$a_c = 3.53$ \\ $b_c = -0.499$} & \thead{$a_e = -0.609$ \\ $b_e = 0.970$} \\
2nd component & \thead{$c_c = 3.477$ \\ $d_c = -0.02$} & \thead{$c_e = -0.609$ \\ $d_e = 0.992$} 
\\
\hline
\end{tabular}
\label{tab:ScalRelCoeffs}
\end{table}

\begin{table*}
\renewcommand{\arraystretch}{1.5}
\centering
\caption{Compilation of the priors used for the fit parameters of the various models. Amplitudes and characteristic frequencies are set according to a power law, as shown at the top of the table. The coefficients for the power law can be found in Table~\ref{tab:ScalRelCoeffs} if not directly specified.}
\begin{threeparttable}
\begin{tabular}{ccccc}
\multicolumn{5}{c}{$f(\textup{constant}, \textup{exponent},\numax) = \textup{constant} + \log_{10}(\numax)^{\textup{exponent}}$} \\
\\
\hline
\multicolumn{1}{c}{\thead{Model specific \\ fit parameters}} & \multicolumn{1}{c}{\thead{Model D}} & \multicolumn{1}{c}{\thead{Model J}} & \multicolumn{1}{c}{\thead{Model H}} & \multicolumn{1}{c}{\thead{Model T}}\\ \hline 
$a$    & \thead{lognormal \\ $\mu=1.2f(a_c, a_e,\numax)$ \\ $\sigma = 0.1$ }  & \thead{lognormal \\ $\mu=f(a_c=3.555, a_e=-1.006,\numax)$ \\ $\sigma = 0.15$ } & \thead{lognormal \\ $\mu=f(a_c, a_e,\numax)$ \\ $\sigma = 0.1$ } & \thead{lognormal \\ $\mu=f(a_c, a_e,\numax)$ \\ $\sigma = 0.1$ }    \\

$b$    & \thead{lognormal \\ $\mu=f(b_c, b_e,\numax)$ \\ $\sigma = 0.3$ }  & \thead{lognormal \\ $\mu=f(b_c, b_e,\numax)$ \\ $\sigma = 0.1$ } & \thead{lognormal \\ $\mu=f(b_c, b_e,\numax)$ \\ $\sigma = 0.1$ } & \thead{lognormal \\ $\mu=f(b_c, b_e,\numax)$ \\ $\sigma = 0.1$ }     \\

$c$    & \thead{--}  & \thead{--} & \thead{lognormal \\ $\mu=f(c_c, c_e,\numax)$ \\ $\sigma = 0.1$ } & \thead{lognormal \\ $\mu=f(c_c, c_e,\numax)$ \\ $\sigma = 0.1$ }     \\

$d$    & \thead{--}  & \thead{lognormal \\ $\mu=f(d_c, d_e,\numax)$ \\ $\sigma = 0.1$ } & \thead{lognormal \\ $\mu=f(d_c, d_e,\numax)$ \\ $\sigma = 0.1$ }& \thead{lognormal \\ $\mu=f(d_c, d_e,\numax)$ \\ $\sigma = 0.1$ } \\

$e$    & \thead{--}  & \thead{--} & \thead{--} & \thead{lognormal \\ $\mu=\frac{1}{3}f(c_c, c_e,\numax)$ \\ $\sigma = 0.5$ } \\

$f$    & \thead{--}  & \thead{--} & \thead{--} & \thead{lognormal \\ $\mu=2.2f(d_c, d_e,\numax)$ \\ $\sigma = 0.5$ }  \\

$l$    & \thead{normal \\ $\mu=2$ \\ $\sigma = 1$}  & \thead{normal \\ $\mu=5/3$ \\ $\sigma = 1$ } & \thead{normal \\ $\mu=4$ \\ $\sigma = 1$ } &\thead{normal \\ $\mu=4$ \\ $\sigma = 1$ }      \\

$k$    & \thead{--}  & \thead{normal \\ $\mu=17/3$ \\ $\sigma = 1$ } & \thead{normal \\ $\mu=4$ \\ $\sigma = 1$ } & \thead{normal \\ $\mu=4$ \\ $\sigma = 1$ }       \\

$m$    & \thead{--}  & \thead{--} & \thead{--} & \thead{normal \\ $\mu=6$ \\ $\sigma = 2$ }  \\
\hline
\multicolumn{5}{c}{\thead{General priors}} \\
\hline
$W$    & \multicolumn{2}{c}{\thead{beta}} & \multicolumn{1}{c}{\thead{$W_\mathrm{est}=\textup{median(Power[-100:])}$ \\ loc=$0.1W_\mathrm{est}$, \\ scale=$10W_\mathrm{est}-$loc}} &  \multicolumn{1}{c}{\thead{$a=1.5$ \\ $b=4$}} \\

$P_\mathrm{osc}$  & \multicolumn{2}{c}{\thead{lognormal}} & \multicolumn{1}{l}{\thead{$\mu=\log_{10}(1e-9)$}} &  \multicolumn{1}{l}{\thead{$\sigma = 0.001$}} \\

$\sigma_\mathrm{osc}$   &  \multicolumn{2}{c}{\thead{lognormal}} & \multicolumn{1}{c}{\thead{$\mu=\log_{10}(10)$}} &  \multicolumn{1}{c}{\thead{$\sigma = 0.001$}} \\

$a_2$    &   \multicolumn{2}{c}{\thead{lognormal}} & \multicolumn{1}{c}{\thead{$\mu=\log_{10}(1e-9)$}} &  \multicolumn{1}{c}{\thead{$\sigma = 0.001$}} \\

$b_2$   &  \multicolumn{2}{c}{\thead{lognormal}} & \multicolumn{1}{c}{\thead{$\mu=\log_{10}(10)$}} &  \multicolumn{1}{c}{\thead{$\sigma = 0.001$}} \\
\hline
\multicolumn{5}{c}{\thead{Variations for KIC 8006161 and the Sun}} \\
\hline
$P_\mathrm{osc}$  & \multicolumn{2}{c}{\thead{lognormal}} & \multicolumn{1}{l}{\thead{$\mu=\log_{10}\left(0.2\cdot\textup{Power}(\sim\numax)\right)$}} &  \multicolumn{1}{l}{\thead{$\sigma =0.1$}} \\  

$\sigma_\mathrm{osc}$    & \multicolumn{2}{c}{\thead{lognormal}} & \multicolumn{1}{l}{\thead{$\mu=\log_{10}\left(2\left(0.26{\numax}^{0.772}\right)\right)$\tnote{*}}} &  \multicolumn{1}{l}{\thead{$\sigma =0.2$}} \\

$a_2$  & \multicolumn{2}{c}{\thead{lognormal}} & \multicolumn{1}{l}{\thead{$P_\mathrm{est} = \textup{median(Power[:10])}$ \\$\mu=\log_{10}\left(\sqrt{P_\mathrm{est}\cdot 10^{b_2}}\right)$}} &  \multicolumn{1}{l}{\thead{$\sigma =0.2$}} \\

$b_2$  & \multicolumn{2}{c}{\thead{lognormal}} & \multicolumn{1}{l}{\thead{$\mu=\log_{10}\left(\textup{Frequency[10])}\right)$}} &  \multicolumn{1}{l}{\thead{$\sigma =0.2$}} \\
\hline

\end{tabular}
\begin{tablenotes}\footnotesize
\item[*] Recommendation for oscillation excess width from \citet{Sreenivas24} based on an approximation of \dnu \citep{Stello09b, Hekker09, Huber11}. 
\end{tablenotes}
\end{threeparttable}
\label{tab:PriorComp}
\end{table*}

\clearpage
\section{Granulation in the Sun}\label{app:C}
\begin{figure}[hb]
    \vspace{5cm}
    \centering
    \begin{minipage}{\textwidth}
        \hspace{0.55cm} 
        \includegraphics[width=17cm]{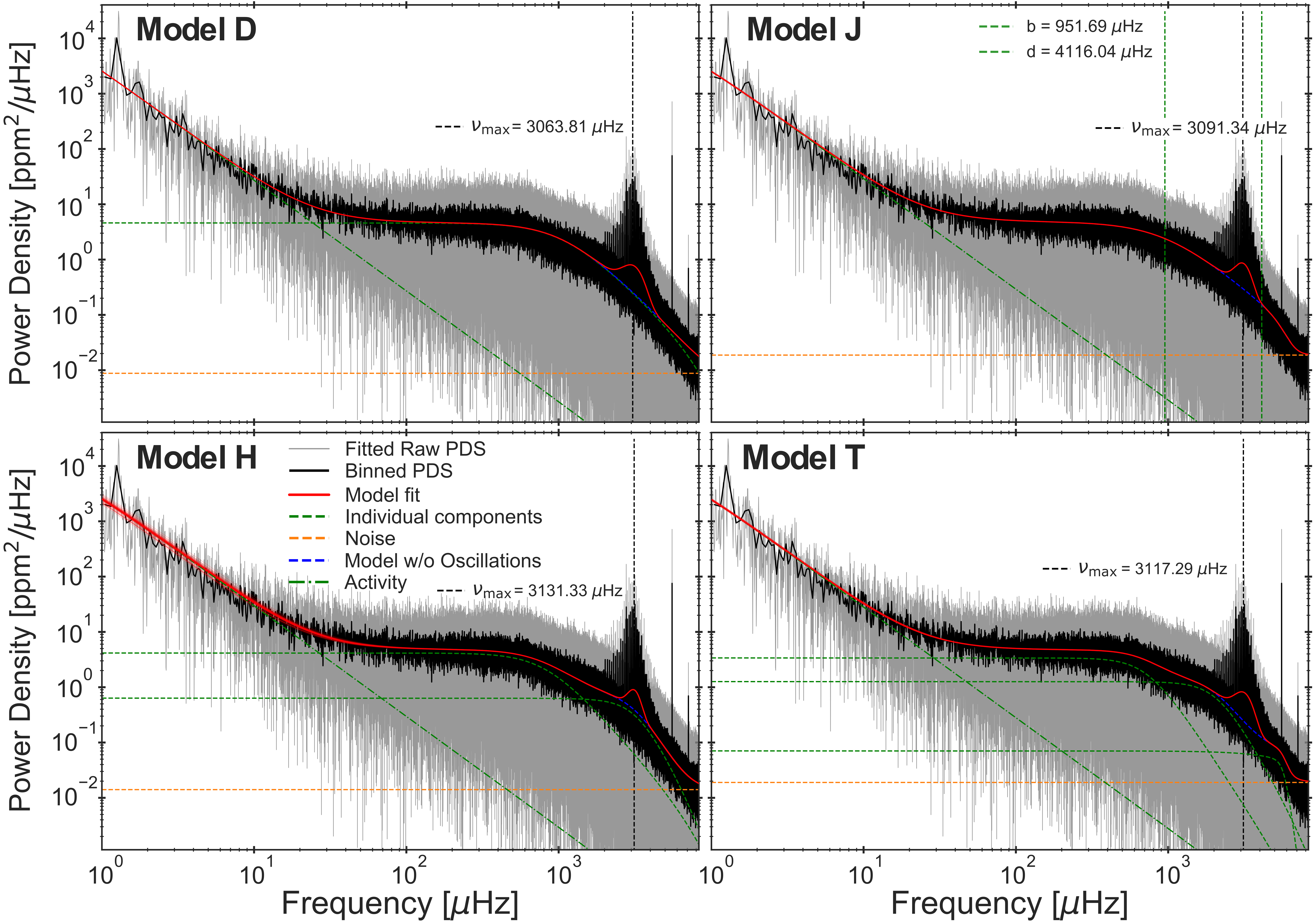}
    \end{minipage}
    \caption{\parbox[t]{1.8\linewidth}{\raggedright Fits of models D, J, H, and T to $\simeq 1150$ days of VIRGO timeseries data of the Sun. The models are fitted to the unbinned PDS shown in grey, but for clarity, a binned version is overplotted. The model is plotted in red using the median of the obtained posteriors for each fit parameter. Additionally, 50 randomly drawn samples from the posteriors are used to repeatedly plot the model alongside, as an indicator of the scatter. For models D, H, and T, the individual components are plotted as green dashed profiles. For model J, instead the values of the two characteristic frequencies are shown as red vertical dashed lines. The fitted value of \numax is given and indicated by the black vertical dashed line, while the noise is shown by the orange horizontal dashed line. The activity component is the green dot-dashed line. The fitted model without the influence of the oscillation excess in plotted as the blue dashed profile.}}
    \label{fig:SunFits}
\end{figure}

As presented in Sect.~\ref{subsec:ValSun}, the developed framework was applied to solar VIRGO data. Figure~\ref{fig:SunFits} shows the results for all four models, while Table~\ref{tab:SunZs} indicate the obtained evidence ratios. The results displayed in the table indicate large preferences for a given model over another, concluding that model T is to be overwhelmingly preferred in the case of the Sun.  

\newpage
\begin{table}[]
\renewcommand{\arraystretch}{1.5}
\centering
\caption{Evidence ratios for model comparison of the Sun.}
\begin{tabular}{cc|cc}
\hline
\thead{Model Comparison} & \thead{$\log(\mathcal{Z}_1)-\log(\mathcal{Z}_2)$} & \thead{Model} & \thead{Ranking} \\
\hline 
D vs. J & $-11145.52 \pm 0.23$ & D & 4\\ 
J vs. H & $2542.57 \pm 0.23$ & J & 2 \\
H vs. T & $-4783.32 \pm 0.23$ & H & 3 \\
J vs. T & $-2240.75 \pm 0.24$ & T & 1\\
\hline
\end{tabular}
\tablefoot{The left-hand columns list the pairwise model comparisons and corresponding evidence ratios. For clarity, the right-hand columns indicate the inferred model and its ranking based on overall support from the evidence. The displayed uncertainties have been estimated by propagating the errors of the individual log-evidences.}
\label{tab:SunZs}
\end{table}
\end{appendix}

\end{document}